\documentclass[lettersize,journal]{IEEEtran}
\usepackage{amsmath,amsfonts}
\usepackage{algorithmic}
\usepackage{algorithm}
\usepackage{array}
\usepackage[caption=false,font=normalsize,labelfont=sf,textfont=sf]{subfig}
\usepackage{textcomp}
\usepackage{stfloats}
\usepackage{multirow}
\usepackage{hyperref}
\usepackage{url}
\usepackage{verbatim}
\usepackage{xspace}
\usepackage{graphicx}
\usepackage{cite}
\usepackage{orcidlink}
\usepackage{booktabs}
\usepackage{makecell}
\makeatletter
\DeclareRobustCommand\onedot{\futurelet\@let@token\@onedot}
\def\@onedot{\ifx\@let@token.\else.\null\fi\xspace}

\def\ie{\emph{i.e}\onedot}

\def\etal{\emph{et al}\onedot}

\makeatother

\renewcommand{\vec}[1]{{\mathbf #1}}
\begin{document}

\title{M2Diff: Multi-Modality Multi-Task Enhanced Diffusion Model for MRI-Guided Low-Dose PET Enhancement}

\author{Ghulam Nabi Ahmad Hassan Yar\orcidlink{0000-0003-0217-1105},
Himashi Peiris\orcidlink{0000-0003-0464-1182},
Victoria Mar\orcidlink{0000-0001-9423-3435}
Cameron Dennis Pain\orcidlink{0000-0003-2729-4259},
Zhaolin Chen\orcidlink{0000-0002-0173-6090}
\thanks{This work does not involve human subjects or animals in its research, and the experiments use public open-access datasets.
}

\thanks{Ghulam Nabi Ahmad Hassan Yar (email: ghulam.yar@monash.edu), Himashi Peiris (email: himashi.peiris@monash.edu), Cameron Dennis Pain (cameron.pain@monash.edu), and Zhaolin Chen (email: zhaolin.chen@monash.edu) are with the Department of DSAI, Faculty of IT, Monash University and Monash Biomedical Imaging, Melbourne, Australia. Victoria Mar (victoria.mar@monash.edu) is with School of Public Health and Preventive Medicine, Monash University and Victorian Melanoma Service, Alfred Health, Melbourne, Australia}
\thanks{© 2026 IEEE. Personal use of this material is permitted. Permission from IEEE must be obtained for all other uses, in any current or future media, including reprinting/republishing this material for advertising or promotional purposes, creating new collective works, for resale or redistribution to servers or lists, or reuse of any copyrighted component of this work in other works.}
\thanks{This is the accepted version of the article: \textit{M2Diff: Multi-Modality Multi-Task Enhanced Diffusion Model for MRI-Guided Low-Dose PET Enhancement}, accepted for publication in IEEE Transactions on Radiation and Plasma Medical Sciences. DOI: \href{https://doi.org/10.1109/TRPMS.2026.3672075}{10.1109/TRPMS.2026.3672075}.}
}

\markboth{Accepted Version at IEEE TRANSACTIONS ON RADIATION AND PLASMA MEDICAL SCIENCES}%
{Shell \MakeLowercase{\textit{et al.}}: A Sample Article Using IEEEtran.cls for IEEE Journals}


\maketitle

\begin{abstract}
Positron emission tomography (PET) scans expose patients to radiation, which can be mitigated by reducing the dose, albeit at the cost of diminished quality. This makes low-dose (LD) PET recovery an active research area. 
Previous studies have focused on standard-dose (SD) PET recovery from LD PET scans and/or multi-modal scans, e.g., PET/CT or PET/MRI, using deep learning.
While these studies incorporate multi-modal information through conditioning in a single-task model, such approaches may limit the capacity to extract modality-specific features, potentially leading to early feature dilution. 
Although recent studies have begun incorporating pathology-rich data, challenges remain in effectively leveraging multi-modality inputs for reconstructing diverse features, particularly in heterogeneous patient populations.
To address these limitations, we introduce a multi-modality multi-task diffusion model (M2Diff) that processes MRI and LD PET scans separately to learn modality-specific features and fuse them via hierarchical feature fusion to reconstruct SD PET.
This design enables effective integration of complementary structural and functional information, leading to improved reconstruction fidelity.
We have validated the effectiveness of our model on both healthy and Alzheimer's disease brain datasets. The M2Diff achieves superior qualitative and quantitative performance on both datasets.
The code is available at: \href{https://github.com/Gnahy/M2Diff}{M2Diff.}
\end{abstract}


\section{Introduction}
\label{sec:intro}

Positron Emission Tomography (PET) is a widely used imaging modality in oncology, neurology, and cardiology. It provides information on the physiological processes of the body to detect and monitor diseases~\cite{LAMEKA2016209}. Its primary advantage lies in its ability to reveal metabolic changes before structural abnormalities appear~\cite{michaelpet}. 
However, PET scans require the use of ionising radiation, limiting the feasibility of repeated follow-ups and applications to pediatric patients. Reducing radiation dose to mitigate these drawbacks introduces higher noise and poorer image quality, which can potentially impact diagnostic accuracy. Methods for recovering standard-dose (SD) PET image quality from low-dose (LD) PET signals aim to reduce radiation exposure as much as possible without compromising diagnostic quality~\cite{HFResDiff}.

Recent advances in deep learning models have revolutionised the domain of medical imaging due to their capability to generate high-quality images. Bousse \etal~\cite{bousse2024review} provided a comprehensive review of neural network–based post-reconstruction denoising techniques for emission tomography, highlighting the potential of deep learning to mitigate photon noise and enhance the visual and quantitative fidelity of low-dose PET and SPECT images. These methods have been used in LD-to-SD PET image recovery in the past few years~\cite{chen2020generalization, zhao2020study, pain2024low, sudarshan2021towards, pain2022deep}. Several studies have used UNet~\cite{chen2020generalization}, GANs~\cite{zhao2020study} and more advanced methods, including DDPM~\cite{Han2023Contrastive, Pan2024Full, xie2024dose} and IDDPM~\cite{yu2024pet} to recover better SD PET image quality. Jiang \etal~\cite{jiang2023pet} utilised latent diffusion models~\cite{Rombach_2022_CVPR} and replaced Gaussian noise with Poisson noise to enable the diffusion model to better adapt to noise in PET images. Xie \etal~\cite{xie2023unified} introduced a Unified Noise-Aware Network (UNN) capable of adapting to varying noise levels across low-count PET acquisitions by combining subnetworks with different denoising strengths, achieving consistent performance across multi-center datasets and outperforming single-noise-level models. Cui \etal~\cite{cui2024mcad} suggested that patient clinical tabular data could facilitate diffusion models in generating higher-quality SD PET images. They also incorporated a discriminator network to further improve the quality of the generated SD PET images. Several other PET/CT and PET/MR-based studies have used multi-modality data for LD-to-SD PET recovery.
Sun \etal~\cite{sun2022high} proposed a GAN-based bi-task architecture with a parallel pipeline that supported the main pipeline for generating high-quality PET images. Their proposed parallel pipeline incorporated T1-weighted images to generate SD PET images, which were then used to compute a bias loss in order to improve the primary pipeline. Jang \etal~\cite{jang2025cross} proposed a Cross-Modality Transformer Network that integrates MR and PET information using spatial and channel-wise self-attention to improve low-dose tau PET image quality, demonstrating superior performance over conventional concatenation-based fusion methods in 18F-MK-6240 tau PET datasets. Xie \etal~\cite{xie2021anatomically} used multi-modality information in a multi-encoder UNet architecture to extract modality-specific features and improve LD-to-SD PET reconstruction. Wang \etal~\cite{wang2023low} evaluated the performance of five state-of-the-art models, including UNet~\cite{ronneberger2015u}, enhanced deep super-resolution network (EDSR)~\cite{lim2017enhanced}, GAN~\cite{goodfellow2014generative}, Swin Transformer~\cite{liang2021swinir}, and EDSR-ViT (vision Transformer)~\cite{carion2020end} for LD-to-SD image restoration using PET/MR data. Several studies have also explored the use of multi-modality data to achieve this task~\cite{xie2025dose} and other tasks, i.e., motion correction~\cite {chen2021mr}, showing the effectiveness of using multi-modality information.

While diffusion-based frameworks have shown promise for PET image recovery, there are several limitations: 
\textbf{(1) }Diffusion models are prone to
underestimating voxel-wise intensities\cite{lin2024common} and causing blurriness \cite{dayarathna2025mccad}. This may cause bias in quantitative PET imaging and result in the loss of high-frequency, clinically significant features, potentially compromising the diagnostic quality of PET scans \cite{HFResDiff}. 
\textbf{(2) }Pathology-rich datasets introduce high-variability, leading to structural diversity that can be difficult to capture~\cite{NEURIPS2021_49ad23d1}, which may result in degraded performance.

To address the above limitations, we propose a multi-modality, multi-task IDDPM (M2Diff) for LD-to-SD PET recovery that explicitly incorporates T1-weighted MRI for structural guidance, prevents feature dilution through a multi-task model, and leverages IDDPM to learn high-variability features in the data.
Our model uses dedicated pathways for T1-weighted MRI and LD PET to learn anatomical and intensity-related information, respectively. It also facilitates capturing distinct features from each modality and preserves modality-specific information to avoid feature dilution~\cite{Marinov_2023_ICCV}. 
To preserve modality-specific representations and improve cross-modal integration, we introduce a Hierarchical Feature Fusion (HFF) strategy that progressively merges features from both streams at multiple decoder stages. This fusion enables the model to leverage both local and global interactions between structural and functional features, enhancing the fidelity of fine-grained reconstructions. 
Crucially, by integrating anatomical priors through MRI and maintaining a task-specific decoding strategy, our model ensures that diagnostically relevant features, such as hypometabolic regions or asymmetric uptake patterns, are faithfully preserved in the recovered PET scans, especially in severe cases where the MRI scan also shows disease biomarkers.
Additionally, IDDPM's stochastic denoising process is inherently more robust to high-variability datasets compared to image-to-image translation models, such as CycleGAN~\cite{Zhu_2017_ICCV}, making it better suited for LD-to-SD PET recovery with pathological variations~\cite{NEURIPS2021_49ad23d1}.

Our contributions include: \textbf{(1)} Developing a novel multi-task model within an IDDPM framework, enabling more effective utilisation of T1-weighted MRI features and capturing diverse pathological variations to enhance PET recovery across different conditions. \textbf{(2)} Integrating T1-weighted MRI images into a diffusion-based framework, leveraging structural brain information to improve LD-to-SD PET recovery. \textbf{(3)} Introducing hierarchical feature fusion and sharing in our multi-task model, enabling layered feature fusion during decoding. This ensures better preservation of structural details to improve the quality and anatomical accuracy of the recovered SD PET. Finally, we evaluate our model on both healthy and Alzheimer’s disease (\ie, ADNI) datasets, which exhibit increased distribution variability due to pathological differences.
\section{Methodology}
\label{sec:method}

\subsection{Problem Formulation}
Throughout the paper, we represent vectors and matrices using bold lower-case letters $\vec{x}$ and bold upper-case letters $\vec{X}$, respectively.
Here, we briefly summarise our notation. Let $\mathcal{X} = \{(\vec{X}_i, \vec{Y}_i, \vec{Z}_i)\}_{i=1}^n$ be a dataset with $n$ samples, where each sample $(\vec{X}_i, \vec{Y}_i, \vec{Z}_i)$ consists of a LD PET scan $\vec{X}_i \in \mathbb{R}^{C \times H \times W}$, its corresponding ground-truth or SD PET scan $\vec{Y}_i \in \mathbb{R}^{C \times H \times W}$, and a T1-weighted MRI scan $\vec{Z}_i \in \mathbb{R}^{C \times H \times W}$. Here, $C$, $H$, and $W$ represent the number of channels, height, and width of the input medical scan, respectively. 
The primary objective is to learn two mapping functions, $\mathcal{G}_1$ and $\mathcal{G}_2$, which together reconstruct the SD PET image and can be viewed as separate decompositions of the overall generator $\mathcal{G}$. Through the hierarchical feature fusion module, both $\mathcal{G}_1$ and $\mathcal{G}_2$ are separately conditioned on the multimodal inputs $\vec{X}_i$ and $\vec{Z}_i$, as well as the diffusion timestep $t$, enabling the model to leverage complementary dose-related and anatomical cues from the dataset $\mathcal{X}$.
For simplicity, the aggregated mapping function is defined with $\vec{\Theta}$ that encapsulates all the parameters of the network associated with two mapping functions (\ie, $\theta_1$, $\theta_2$):
\begin{equation}
    \hat{\vec{Y}}_{i} = \mathcal{G}(\vec{\Theta}; \vec{Y}_{t}, \vec{X}_{i}, \vec{Z}_{i}, t),
\end{equation}
where 
$t \in \{1, \dots, T\}$ represents the discrete time step in the reverse diffusion process.

\subsection{Improved Denoising Diffusion Probabilistic Model (IDDPM)}

We use conditional IDDPM~\cite{pmlr-v139-nichol21a} as the task-specific backbone model in our multi-task framework. IDDPM consists of two main parts: the forward process and the reverse process. In the forward process, Gaussian noise $\epsilon$ is iteratively added to SD PET ($\vec{Y}_{0}$) over T steps, gradually transforming it into pure noise ($\vec{Y}_{T}$). While the reverse process reconstructs $\vec{Y}_{0}$ by iteratively denoising $\vec{Y}_{t}$ guided by a conditioning mechanism. The primary difference between DDPM and IDDPM lies in how noise is modelled. While DDPM uses a learned noise mean $\mu_{\theta}$ with a fixed variance schedule, IDDPM introduces a learned variance $\Sigma_{\theta}$ along with the mean, allowing for more flexible noise estimation and improved sample quality. Inspired by IDDPM~\cite{pmlr-v139-nichol21a}, the reverse process can be defined as:
\begin{equation}
    p_{\theta}(\vec{Y}_{0}|\vec{Y}_{t}, \vec{\Lambda})  =  \mathcal{N}\big(\vec{Y}_{t-1}; \mu_{\theta}(\vec{Y}_{t}, t; \vec{\Lambda}), \Sigma_{\theta}(\vec{Y}_{t}, t; \vec{\Lambda})\big),
    \label{eqn:iddpm}
\end{equation}
where $\theta$ denotes the set of learnable parameters of the neural network used to approximate the reverse diffusion distribution, and $\vec{\Lambda}$ represents the conditioning mechanism, which includes $\vec{X}_{i}$ and $\vec{Z}_{i}$. We parameterise $\mu_{\theta}(\vec{Y}_{t}, t;\vec{\Lambda})$, and $\Sigma_{\theta}(\vec{Y}_{t}, t;\vec{\Lambda})$ to directly predict $\vec{Y}_{0}$ at each timestep $t$, and controlled noise $\epsilon_{t-1}$ is reintroduced to generate the next noisy estimate $\vec{Y}_{t-1}$.

\subsection{Conditional Learning in M2Diff}
Unlike the original IDDPM, which predicts the noise component $\epsilon_t$, 
our model directly predicts the denoised SD-PET image $\hat{\vec{Y}}_{0}$ at every diffusion step.
At each iteration, a timestep $t$ is uniformly sampled as $t \sim \mathcal{U}(1,T)$,
and Gaussian noise $\epsilon_t \sim \mathcal{N}(0,I)$ is added to the clean SD-PET image $\vec{Y}_0$
to obtain its noisy counterpart $\vec{Y}_t$.
The network receives $(\vec{Y}_t, \vec{X}_i, \vec{Z}_i, t)$ as input and predicts $\hat{\vec{Y}}_{0}$,
which serves as an estimate of the fully denoised image at that timestep.
The next step’s sample $\vec{Y}_{t-1}$ is generated by re-adding controlled noise.

\begin{figure*}
\centering
\includegraphics[width=\textwidth]{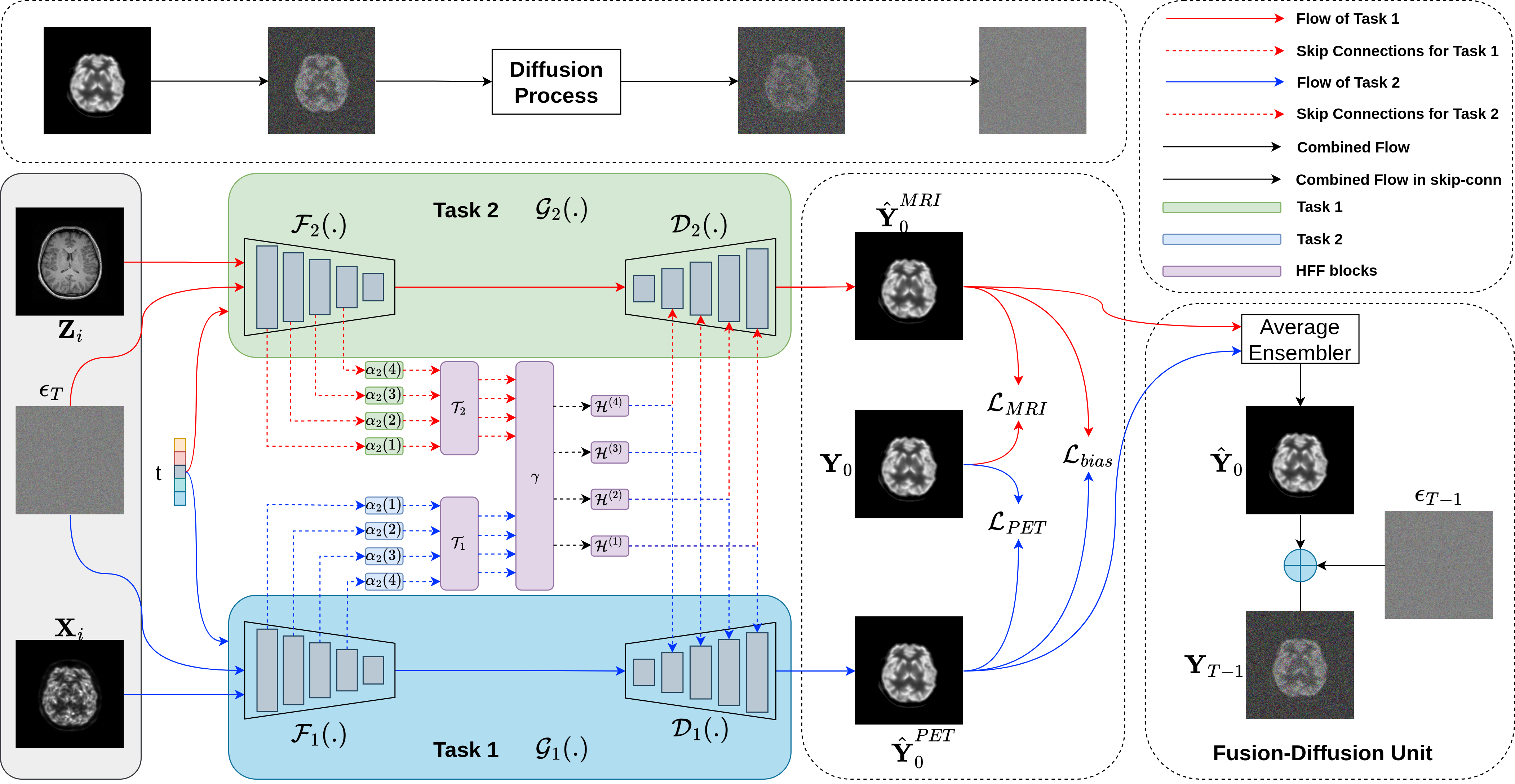}
\caption{Overview of the proposed \textbf{M2Diff} architecture for multi-task PET reconstruction. The model takes low-dose PET ($\mathbf{X}_i$) and anatomical MRI ($\mathbf{Z}_i$) as inputs and comprises two task-specific branches—\textbf{Task 1} and \textbf{Task 2}—targeting complementary PET reconstruction goals. Each branch includes an encoder $\mathcal{F}_k$, and a decoder $\mathcal{D}_k$ ($k \in {1, 2}$). Features from all encoder levels are fused through a \textbf{Hierarchical Feature Fusion} module, enabling cross-task interaction. Final outputs $\hat{\mathbf{Y}}_0$ are optimised via hybrid losses.}
\label{fig_pipeline}
\end{figure*}

The timestep $t$ is encoded via a sinusoidal embedding~\cite{vaswani2017attention} $E_t$ that provides temporal context about the current noise level.
$E_t$ is injected into both encoders and decoders of M2Diff, modulating convolutional and attention blocks as shown in Fig.~\ref{fig_pipeline}.
This conditioning allows the network to learn how much detail to restore at each step.

Following the IDDPM formulation~\cite{pmlr-v139-nichol21a},
the network predicts both the mean $\mu_\theta$ and variance $\Sigma_\theta$ of the reverse process.
In our dual-decoder setup, $\Sigma_{\theta_1}$ and $\Sigma_{\theta_2}$ are learned separately for the PET- and MRI-conditioned branches.
These variance estimates act as uncertainty indicators; higher $\Sigma_\theta$ corresponds to regions of greater noise or anatomical ambiguity,
helping stabilise sampling and improving fidelity in challenging regions such as low-uptake or motion-affected areas.

\subsection{M2Diff Architecture}

An overview of M2Diff is presented in Fig.~\ref{fig_pipeline}. M2Diff introduces a novel multi-modality, multi-task diffusion model for LD-to-SD PET recovery, leveraging hierarchical fusion and task-specific encoder-decoder architecture to enhance reconstruction quality.
Unlike a single-task model where multiple conditions are applied without explicit separation, causing early feature dilution, our model employs different paths for extracting intensity and structural-related information, ensuring that modality-specific features are preserved.

Each generation task (or mapping function) $\mathcal{G}_1(\cdot)$ and $\mathcal{G}_2(\cdot)$ comprises an encoder $\mathcal{F}_1(\cdot)$ and $\mathcal{F}_2(\cdot)$, and a decoder $\mathcal{D}_1(\cdot)$ and $\mathcal{D}_2(\cdot)$, where subscript 1 and 2 represents task 1 for LD-PET to SD-PET and task 2 for T1-weighted MRI to SD-PET image generation, respectively. The encoder and decoder network for both tasks was inspired by IDDPM work by Nichol and Dhariwal~\cite{pmlr-v139-nichol21a}. Inspired by the work of Le \etal~\cite{li2020hierarchical}, we used
the hierarchical feature fusion (HFF) to facilitate the sharing of complementary information from both encoders at the decoder stage. Finally, a dual-decoder architecture ensures that each pathway contributes separately to SD PET recovery before applying average ensembling to get a final prediction. Each decoder follows a similar architectural design; this principle is widely explored in segmentation tasks to enhance mutual-consistency learning~\cite{wu2021semi,wu2022mutual}, allowing each decoder to capture modality-specific contextual information while maintaining coherent reconstruction across modalities.

These design choices make M2Diff robust to feature dilution and improve generalizability, even in datasets with high pathological variability.

\textit{Multi-task Learning} is the core of our model that integrates multiple image modalities, including $\vec{Z}_{i}$ and $\vec{X}_{i}$. The first encoder \textbf{$\mathcal{F}_{1}$} uses $\vec{X}_{i}$ as a conditioning mechanism to extract intensity-related information, while the second encoder \textbf{$\mathcal{F}_{2}$} uses $\vec{Z}_{i}$ as a conditioning mechanism to extract anatomical information of the brain. This dual-encoder design ensures that modality-specific information is preserved by preventing early feature dilution among different modalities. This enhances the models' ability to extract complementary information, resulting in a more precise reconstruction.

$\mathcal{F}_{1}$ and $\mathcal{F}_{2}$ have a similar architecture that utilises convolutional layers, residual connections, and attention mechanisms to capture modality-specific features effectively. Each $\mathcal{F}$ extracts different features,

\begin{align}
    \vec{\alpha}_{1}{(l)} &= \mathcal{F}_{1}(\theta_{1}; \vec{X}_i, \vec{Y}_t, t)[l], \quad l = 1, \dots, L, \notag \\
    \vec{\alpha}_{2}{(l)} &= \mathcal{F}_{2}(\theta_{2}; \vec{Z}_i, \vec{Y}_t, t)[l], \quad l = 1, \dots, L.
\end{align}

Here, $\vec{\alpha_{1}(l)}$ and $\vec{\alpha_{2}(l)}$ represent feature maps extracted at layer \textit{l} from modality specific encoder $\mathcal{F}_{1}$ and $\mathcal{F}_{2}$, respectively and $l \in \{1, 2, \dots, L\}$ denotes the layer index in the encoder network excluding encoder output layer, where $L+1$ is the total number of layers and layer $L+1$ denotes the output of the encoder. This structured encoding allows feature extraction from each modality before fusion, unlike direct concatenation-based conditioning in single-task models. These features are fused hierarchically before passing them to $\mathcal{D}_{1}$ and $\mathcal{D}_{2}$.

To exploit cross-modal complementary information from both encoders, we adopt a hierarchical feature integration strategy that merges representations at each layer. For each layer $l$, the features \(\vec{\alpha}_1{(l)}\) and \(\vec{\alpha}_2{(l)}\) are first projected into a shared representational space via learned linear transformations \(\mathcal{T}_1\) and \(\mathcal{T}_2\), and then integrated through a composition function that operates along the channel axis:
\begin{equation}
\mathcal{H}^{(l)} = \gamma\Big(  \mathcal{T}_1\big( \vec{\alpha}_1{(l)} \big) \,\|\, \mathcal{T}_2\big( \vec{\alpha}_2{(l)} \big)  \Big), \quad l = 1, \dots, L.
\end{equation}
Here, \(\|\) denotes channel-wise concatenation, and \(\gamma\) is a non-linear integration head that refines and reduces the dimensionality of the fused representations.
Collectively, the multi-level fused representations are denoted as:
\begin{equation}
    \mathcal{H}_{\text{all}} = \{\mathcal{H}^{(1)}, \mathcal{H}^{(2)}, \dots, \mathcal{H}^{(L)}\}.
\end{equation}

In summary, this operation yields a hierarchy of semantically enriched feature maps $\mathcal{H}_{\text{all}}$, which encapsulate complementary spatial and contextual information across modalities. The fused representation is used as a conditioning pathway for each decoder $\mathcal{D}_{1}$ and $\mathcal{D}_{2}$, which separately reconstructs the SD-PET image from its corresponding modality input via:
\begin{equation}
    \mathcal{G}_1(\theta_1; \vec{X}_{i}, \vec{Y}_{t}, t) = \mathcal{D}_{1}\Big(\theta_1; \mathcal{H}_{\text{all}}, \mathcal{F}_{1}\big(\theta_1; \vec{X}_i,\vec{Y}_{t}, t\big)[L+1] \Big),
\end{equation}
\begin{equation}
    \mathcal{G}_2(\theta_2; \vec{Z}_{i}, \vec{Y}_{t},t) = \mathcal{D}_{2}\Big(\theta_2; \mathcal{H}_{\text{all}}, \mathcal{F}_{2}\big(\theta_2; \vec{Z}_i,\vec{Y}_{t}, t \big)[L+1] \Big).
\end{equation}

This dual-decoder design ensures that both intensity and structural pathways separately contribute to SD PET recovery rather than competing within a single network. Each encoder specialises in reconstruction $\vec{Y}_{0}$ based on its source modality and shared contextual information via $\mathcal{H}_{\text{all}}$. The final output is computed using the average ensembling of the two modality-specific predictions,
\begin{equation}
    \hat{\vec{Y}} = \frac{1}{2} \big(\hat{\vec{Y}}_{0}^{\text{PET}} + \hat{\vec{Y}}_{0}^{\text{MRI}}\big).
\end{equation}

Here, $\hat{\vec{Y}}_{0}^{\text{MRI}}$ and $\hat{\vec{Y}}_{0}^{\text{PET}}$ represent the SD-PET output generated by MRI and PET-driven decoding branches, respectively. The averaging promotes robustness and balances modality-specific biases, leading to improved reconstruction fidelity.

\subsection{Objective Function}

The objective function used for training consists of two parts: 1) image recovery loss from the prediction of both decoders $\mathcal{L}_{\text{PET}}$ and $\mathcal{L}_{\text{MRI}}$, and 2) a modality consistency regularization term or bias loss ($\mathcal{L}_{\text{bias}}$). Specifically, $\mathcal{L}_{\text{PET}}$ and $\mathcal{L}_{\text{MRI}}$ are Mean Squared Error (MSE) calculated between $\vec{Y}_{0}$ and their respective output.
\begin{align}
    \mathcal{L}_{\text{PET}}(\theta_1;\mathcal{X}) = \text{MSE}(\hat{\vec{Y}}_{0}^{\text{PET}}, \vec{Y}_0), \\
    \mathcal{L}_{\text{MRI}}(\theta_2;\mathcal{X}) = \text{MSE}(\hat{\vec{Y}}_{0}^{\text{MRI}}, \vec{Y}_0).
\end{align}

Inspired by prior studies in medical imaging tasks with a multi-view setting~\cite{peiris2023uncertainty, sun2022high}, to encourage consistency between the predictions from both modalities, we introduce a bias regularisation loss:
\begin{equation}
    \mathcal{L}_{\text{bias}}(\Theta;\mathcal{X}) = \text{MSE}(\hat{\vec{Y}}_{0}^{\text{PET}}, \hat{\vec{Y}}_{0}^{\text{MRI}}),
\end{equation}

The overall multi-task loss function is defined as:
\begin{equation}
    \mathcal{L}(\Theta, \mathcal{X}) = \lambda_{1}(\mathcal{L}_{\text{PET}} + \mathcal{L}_{\text{MRI}}) + \lambda_{2} \mathcal{L}_{\text{bias}},
    \label{eq_loss}
\end{equation}
where $\lambda_{1}$ and $\lambda_{2}$ are weights that determine the contribution of restoration loss from both task-specific models and bias loss, respectively.

\section{Experiments}
\label{sec:experiments}

\subsection{Experimental Setup}

\begin{figure*}[h!]
\centering
\includegraphics[width=0.9\textwidth]{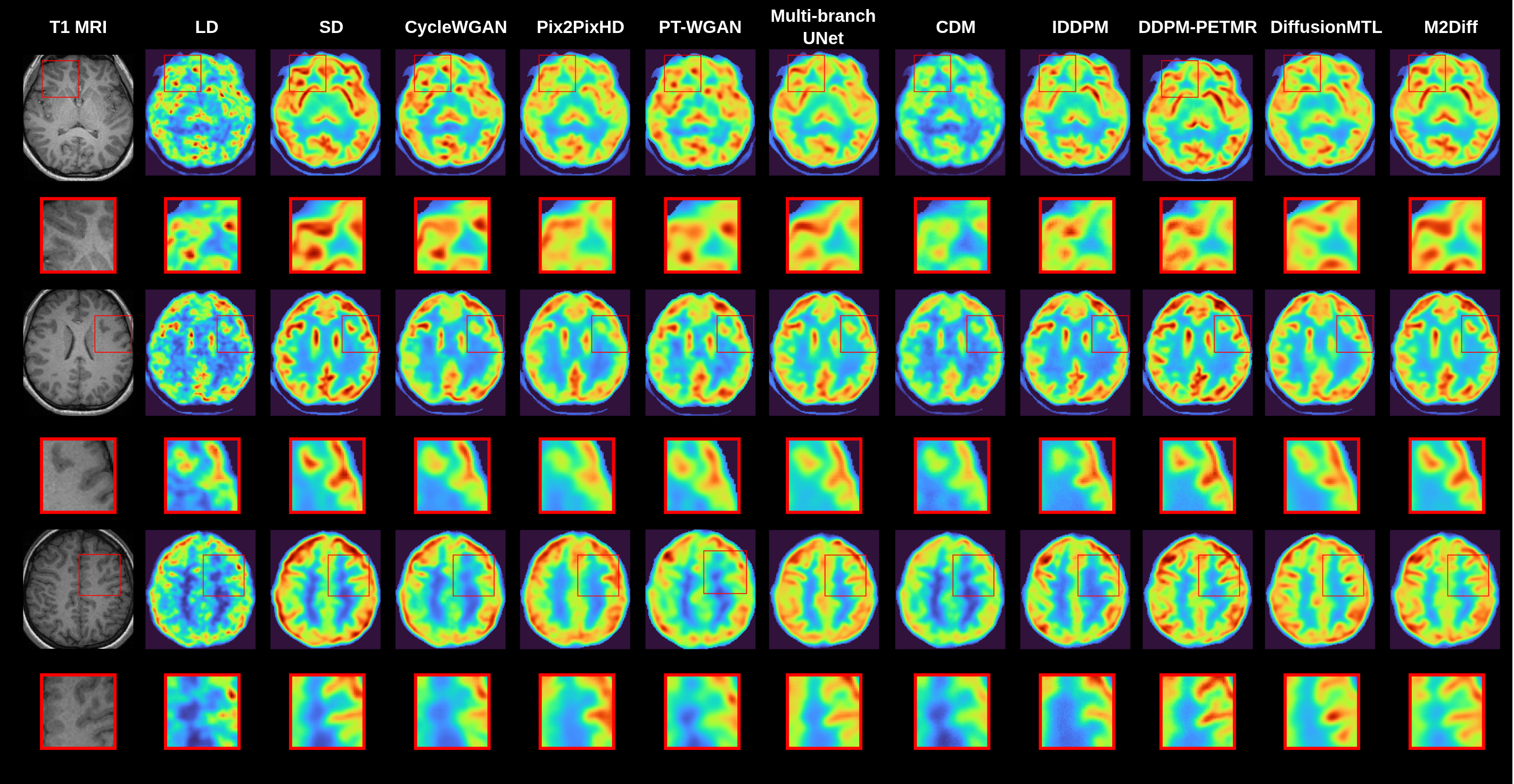}
\caption{Comparison of reconstructed PET images across baseline models and the proposed M2Diff on the DaCRA dataset with DRF of $\times$100. Each row corresponds to a different subject scan, and each column represents a different reconstruction method, including T1-weighted MRI, low-dose input (LD), standard-dose ground truth (SD), and competing methods: CycleWGAN, Pix2PixHD, PT-WGAN, Multi-branch UNet, CDM, IDDPM, DDPM-PETMR, DiffusionMTL, and our proposed M2Diff. The upper portion of each tile shows the full axial brain slice, while the lower zoomed-in views (red-bordered) highlight regions of interest. All PET images, including zoomed ROIs, are displayed using a fixed intensity range of [0,1] for all methods.} \label{fig_baseline}
\end{figure*}

\begin{figure*}[h!]
\centering
\includegraphics[width=0.9\textwidth]{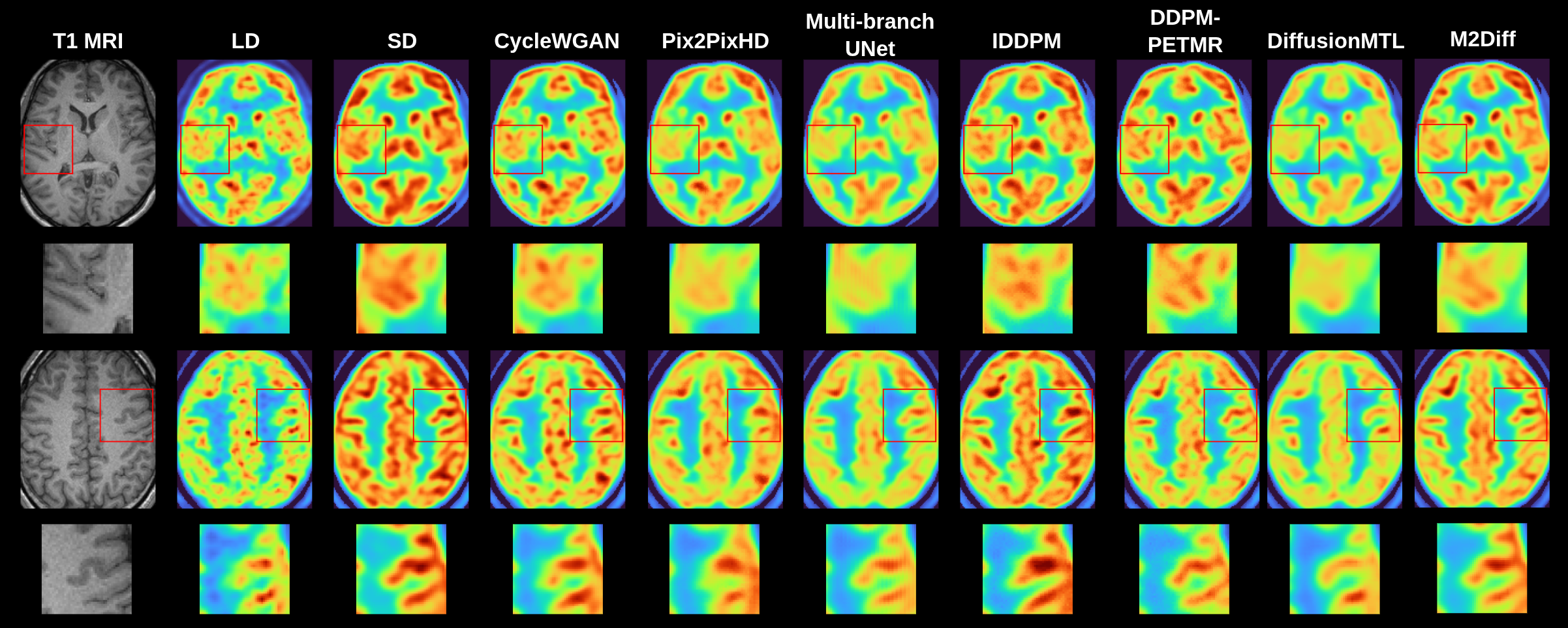}
\caption{Comparison of reconstructed PET images across baseline models and the proposed M2Diff on the DaCRA dataset with DRF of $\times$20. Each row corresponds to a different subject scan, and each column represents a different reconstruction method, including T1-weighted MRI, low-dose input (LD), standard-dose ground truth (SD), and competing methods: CycleWGAN, Pix2PixHD, Multi-branch UNet, IDDPM, DDPM-PETMR, DiffusionMTL, and our proposed M2Diff. The upper portion of each tile shows the full axial brain slice, while the lower zoomed-in views (red-bordered) highlight regions of interest. All PET images, including zoomed ROIs, are displayed using a fixed intensity range of [0,1] for all methods.} \label{fig_baseline_x20}
\end{figure*}

\begin{table*}[h!]
\caption{Quantitative results of comparative study on DaCRA Dataset with $\times$100 DRF}\label{table_baseline}
\centering
\resizebox{0.65\linewidth}{!}{%
\begin{tabular}{l l l l l l}
\toprule
& \multicolumn{1}{c}{\textbf{Model}} 
& \multicolumn{1}{c}{\textbf{SSIM $\uparrow$}} 
& \multicolumn{1}{c}{\textbf{PSNR $\uparrow$}} 
& \multicolumn{1}{c}{\textbf{NMSE $\downarrow$}} 
& \multicolumn{1}{c}{\textbf{LPIPS $\downarrow$}} \\
\midrule
\multirow{3}{*}{\makecell{Non-Diffusion \\ Methods}} & CycleWGAN~\cite{zhou2020supervised}     & 0.9427$\pm$\scriptsize{0.0370} & 28.1179$\pm$\scriptsize{2.4479}          & \textbf{0.0621$\pm$\scriptsize{0.0799}} & 0.0374$\pm$\scriptsize{0.0331} \\ 
& Pix2PixHD~\cite{hosch2022artificial}     & 0.9381$\pm$\scriptsize{0.0363}          & 28.1343$\pm$\scriptsize{2.4417} & 0.0695$\pm$\scriptsize{0.1167}          & 0.0423$\pm$\scriptsize{0.0293}          \\ 
& PT-WGAN~\cite{gong2020parameter}         & 0.9304$\pm$\scriptsize{0.0421}          & 28.1800$\pm$\scriptsize{1.7527}          & \underline{0.0630$\pm$\scriptsize{0.1758}} & 0.0496$\pm$\scriptsize{0.0172} \\ 
& Multi-Branch UNet~\cite{xie2021anatomically}         & \underline{0.9498$\pm$\scriptsize{0.0343}}          & \underline{28.4957$\pm$\scriptsize{2.4580}}          & 0.0648$\pm$\scriptsize{0.0947} & \underline{0.0370$\pm$\scriptsize{0.0290}}         \\ \midrule
\multirow{4}{*}{\makecell{Diffusion-based \\ Methods}} & CDM~\cite{Han2023Contrastive} & 0.9275$\pm$\scriptsize{0.0463}          & 25.8563$\pm$\scriptsize{3.3323}          & 0.0958$\pm$\scriptsize{0.0859}          & 0.0517$\pm$\scriptsize{0.0380}          \\ 
& IDDPM~\cite{yu2025robust}                   & 0.9340$\pm$\scriptsize{0.0414}          & 27.1968$\pm$\scriptsize{2.3031}          & 0.0751$\pm$\scriptsize{0.0939}          & 0.0434$\pm$\scriptsize{0.0407}          \\ 
& DDPM-PETMR~\cite{gong2024pet}                  & 0.9462$\pm$\scriptsize0.0335                          & 28.1048$\pm$\scriptsize2.5155                          & 0.0646$\pm$\scriptsize0.0766 & 
0.0378$\pm$\scriptsize0.0277          \\ 
& DiffusionMTL~\cite{ye2024diffusionmtl}   & 0.9267$\pm$\scriptsize{0.0386}          & 26.5145$\pm$\scriptsize{2.2397}          & 0.0931$\pm$\scriptsize{0.1183}          & 0.0481$\pm$\scriptsize{0.0359}          \\ 
& M2Diff (Ours)                                  & \textbf{0.9528$\pm$\scriptsize{0.0308}} & \textbf{28.6369$\pm$\scriptsize{2.7353}} & 0.0694$\pm$\scriptsize{0.1182}          & \textbf{0.0349$\pm$\scriptsize{0.0298}} \\ \bottomrule
\end{tabular}
}
\end{table*}

\begin{table*}[h!]
\caption{Quantitative results of comparative study on DaCRA Dataset with $\times$20 DRF}\label{table_baseline_x20}
\centering
\resizebox{0.65\linewidth}{!}{%
\begin{tabular}{l l l l l l}
\toprule
& \multicolumn{1}{c}{\textbf{Model}} 
& \multicolumn{1}{c}{\textbf{SSIM $\uparrow$}} 
& \multicolumn{1}{c}{\textbf{PSNR $\uparrow$}} 
& \multicolumn{1}{c}{\textbf{NMSE $\downarrow$}} 
& \multicolumn{1}{c}{\textbf{LPIPS $\downarrow$}} \\
\midrule
\multirow{3}{*}{\makecell{Non-Diffusion \\ Methods}} & CycleWGAN~\cite{zhou2020supervised}     &  0.9653$\pm$\scriptsize{0.0123}& 30.8462$\pm$\scriptsize{2.2905}          & 0.0212$\pm$\scriptsize{0.0270} & 0.0195$\pm$\scriptsize{0.0080} \\ 
& Pix2PixHD~\cite{hosch2022artificial}     &     0.9643$\pm$\scriptsize{0.0114}      & \underline{31.175$\pm$\scriptsize{2.5819}} & 0.0215$\pm$\scriptsize{0.0474}    &   0.0187$\pm$\scriptsize{0.0064}        \\ 
& Multi-Branch UNet~\cite{xie2021anatomically}         &    \textbf{0.9712$\pm$\scriptsize{0.0108}}       &    30.2132$\pm$\scriptsize{2.8871} & 0.0226$\pm$\scriptsize{0.0176} &  \underline{0.0171$\pm$\scriptsize{0.0067}}        \\ \midrule
\multirow{4}{*}{\makecell{Diffusion-based \\ Methods}} &  IDDPM~\cite{yu2025robust}                   &    0.9589$\pm$\scriptsize{0.0128}   &     30.1130$\pm$\scriptsize{2.2446}      & 0.0241$\pm$\scriptsize{0.0289}          &    0.0217$\pm$\scriptsize{0.0091}       \\ 
& DDPM-PETMR~\cite{gong2024pet}    &     0.9625$\pm$\scriptsize{0.0126}     & 30.4225$\pm$\scriptsize{2.0816}      & \underline{0.0208$\pm$\scriptsize{0.0166}} & 0.0203$\pm$\scriptsize{0.0088}
          \\ 
& DiffusionMTL~\cite{ye2024diffusionmtl}    &     0.9595$\pm$\scriptsize{0.0138}     & 28.3701$\pm$\scriptsize{2.5887}      & 0.0316$\pm$\scriptsize{0.0139} & 0.0229$\pm$\scriptsize{0.0086}
          \\ 
& M2Diff (Ours)  & \underline{0.9712$\pm$\scriptsize{0.0114}} & \textbf{31.7793$\pm$\scriptsize{1.7985}} & \textbf{0.0154$\pm$\scriptsize{0.0119}}          & \textbf{0.0160$\pm$\scriptsize{0.0076}}  \\ \bottomrule
\end{tabular}
}
\end{table*}

\paragraph{\textbf{Datasets}} DaCRA~\cite{10.1093/gigascience/giac031} data included twenty-seven subjects, who were administered approximately 250 MBq of $^{18}\textrm{F-FDG}$ using a 3T Siemens Biograph PET-MR system.
The Dose was administered for 30 minutes, and data was collected for 60 minutes. Scatter and attenuation correction were applied using MR-derived $\mu$-maps. PET images were reconstructed using an ordered subsets expectation maximisation algorithm with 3 iterations, 21 subsets, and point spread function modelling via Siemens e7tools.
The dose reduction factor (DRF) of $\times$100 LD data was generated via random sampling of acquired list-mode data using Siemens e7Tools with the same parameters used for reconstructing SD data. T1-weighted images were acquired simultaneously with PET images, and non-linear registration was applied using ANTs symmetric normalisation algorithm~\cite{tustison2021antsx}.

The second experimental dataset was taken from Alzheimer's Disease Neuroimaging Initiative (ADNI)~\cite{jack2008alzheimer}. The ADNI database consisted of PET and MRI scans for three classes: normal, mild cognitive impairment, and Alzheimer's disease. 
We simulated DRF $\times100$ by first forward-projecting the data to the sinogram domain using Software for Tomographic Image Reconstruction (STIR)~\cite{thielemans2012stir}. Random sampling was applied in the sinogram domain, and images were reconstructed using a maximum likelihood expectation maximisation (MLEM). Sinogram resampling was performed such that the peak signal-to-noise ratio of the reconstructed images matched $\times 100$ DRF images from the DaCRA dataset\cite{sudarshan2021towards}.

All reconstructed PET volumes were spatially aligned and intensity-normalized to [0, 1]. For the M2Diff model, we used 2D axial slices extracted from the reconstructed 3D volumes. Slices were uniformly sampled across the entire axial range of each subject to ensure equal representation of all anatomical regions and prevent sampling bias toward high-uptake regions. During training, each slice was normalized to [0, 1]. For qualitative visualization, no additional per-slice or per-ROI normalization was applied to the reconstructed predictions. This strategy avoids bias due to varying uptake magnitudes while preserving relative uptake distributions within each image. Instead, all PET slices were displayed using the same fixed intensity range [0,1].

\paragraph{\textbf{Implementation Details}} Our model was implemented using the PyTorch framework~\cite{paszke2017automatic} and was trained end-to-end on NVIDIA A40 GPU for 100 epochs with Adam optimiser. The learning rate was set to $1e^{-4}$, and diffusion step T was set to 1000. Controllable Weights for multi-task loss function $\lambda_{1}$ and $\lambda_{2}$ were empirically set to 0.4 and 0.2, respectively.

\subsection{Comparative Experiments}

We compared the proposed M2Diff model against a set of \textbf{eight} state-of-the-art image generation baselines on the x100 DRF of DaCRA dataset, including both diffusion-based and non-diffusion-based methods from the LD-to-SD PET synthesis literature. The comparison was based on four quantitative metrics: PSNR, NMSE, SSIM, and LPIPs\footnote{PSNR: Peak Signal-to-Noise Ratio, NMSE: Normalized Mean Squared Error, SSIM: Structural Similarity Index Measure, LPIPS: Learned Perceptual Image Patch Similarity.}. The non-diffusion baselines include 1) Pix2PixHD generator~\cite{hosch2022artificial}; 2) CycleWGAN~\cite{zhou2020supervised}; 3) PT-WGAN~\cite{gong2020parameter}; 4)Multi-branch UNet~\cite{xie2021anatomically} . Diffusion-based methods include: 5) Contrastive Diffusion Model (CDM) with Auxiliary Guidance~\cite{Han2023Contrastive}; 6) IDDPM~\cite{yu2025robust}; {7) DDPM-PETMR~\cite{gong2024pet}  8) DiffusionMTL \cite{ye2024diffusionmtl}. Among the baseline, Multi-branch UNet, DDPM-PETMR, and DiffusionMTL are multi-modality models that takes both PET and MRI as input.} The comparison results for the DaCRA dataset with DRF of $\times100$ are shown in Fig.~\ref{fig_baseline} and Table~\ref{table_baseline}. We also conducted a study on the $\times20$ DRF of DaCRa dataset on top 3 non-diffusion and diffusion baselines and our proposed M2Diff as shown in Fig.~\ref{fig_baseline_x20} and Table~\ref{table_baseline_x20}.

\begin{figure}
\centering
\includegraphics[width=\linewidth]{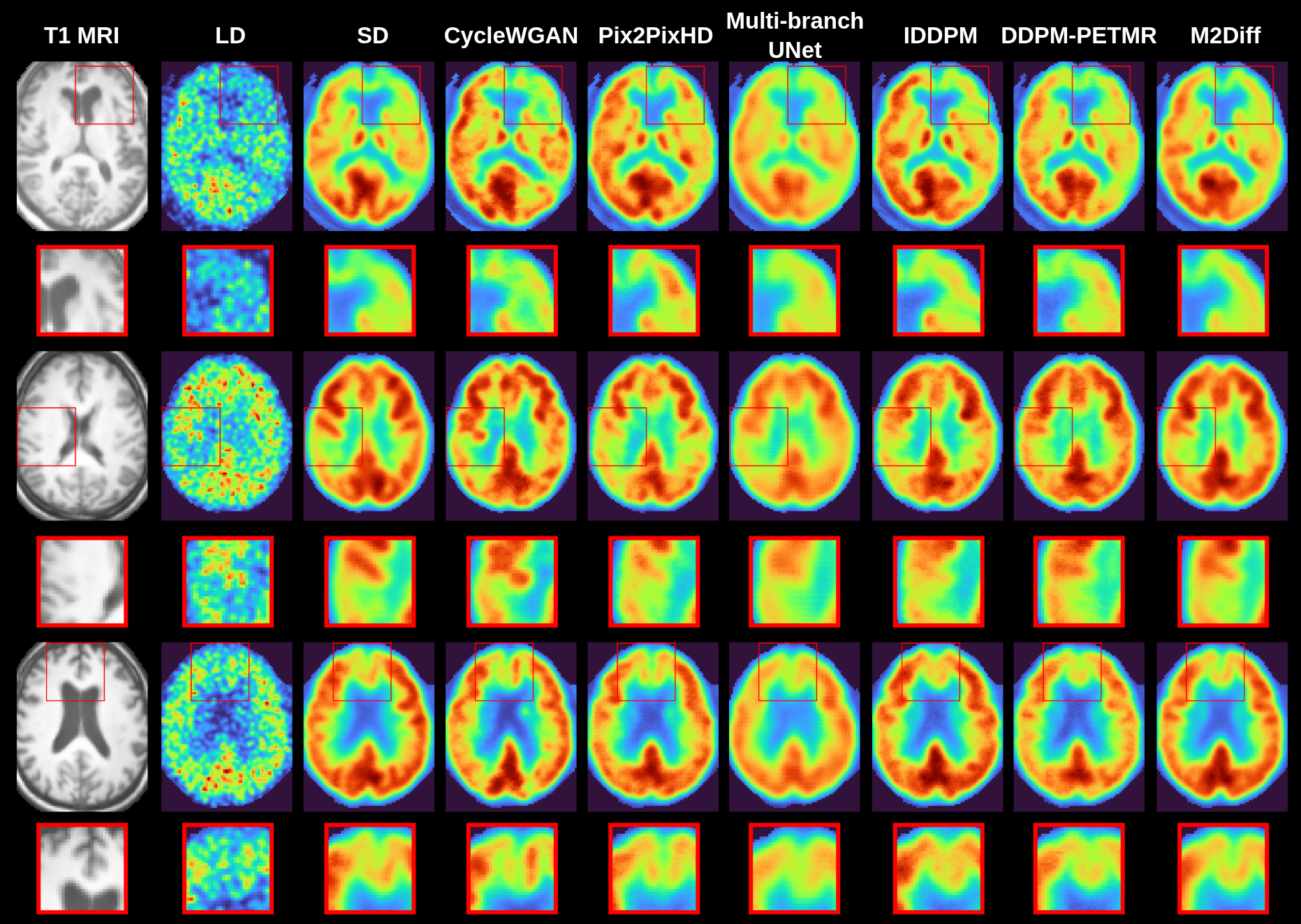}
\caption{Qualitative comparison of reconstructed PET images across baseline models and the proposed M2Diff on the ADNI dataset. Each row corresponds to a different patient scan, and each column represents a different reconstruction method, including T1-weighted MRI, low-dose input (LD), standard-dose ground truth (SD), and competing methods: CycleWGAN, Pix2PixHD, Multi-branch UNet, IDDPM, DDPM-PETMR, and our proposed M2Diff. The upper portion of each tile shows the full axial brain slice, while the lower zoomed-in views (red-bordered) highlight regions of interest. All PET images, including zoomed ROIs, are displayed using a fixed intensity range of [0,1] for all methods.}
\label{fig_ADNI}
\end{figure}

In addition to that, we conducted a comparative study for the best-performing models and M2Diff using the ADNI dataset. The qualitative and quantitative results for this dataset are shown in Fig.~\ref{fig_ADNI} and Table~\ref{table_ADNI}, respectively. In Table~\ref{table_baseline},~\ref{table_baseline_x20} and~\ref{table_ADNI}, the best results are shown in bold, and the second-best results are underlined.

For the DaCRA dataset, M2Diff outperformed other baseline methods based on SSIM, PSNR, and LPIPS, while it performed competitively in the NMSE. Multi-branch UNet achieved the second-best score among all the baselines in the DaCRA dataset, and M2Diff outperformed it in terms of SSIM, PSNR, and LPIPS for $\times100$ DRF. For $\times20$ DRF Multi-branch UNet showed slightly better SSIM but failed to compete with the proposed model in terms of PSNR, NMSE, and LPIPS. 

Qualitatively comparing the results showed that our proposed multi-task model successfully reconstructed the details more effectively compared to all other models for both DRFs. The representative brain regions with changes are shown in Fig.~\ref{fig_baseline} and~\ref{fig_baseline_x20}. The zoomed brain regions in red boxes demonstrated improved accuracy in the recovery of cortical grey matter brain structures. 

For the ADNI dataset, M2Diff further outperformed other models, with IDDPM showing the second-best results. Fig.~\ref{fig_ADNI} shows the qualitative performance of different models on scans from a patient suffering from Alzheimer's disease. Upon examining the PET activities from the generated SD PET scans, a reduction in dose uptake was evident, particularly in the frontal and temporal lobe regions typically affected by Alzheimer’s pathology. Multi-branch UNet and DDPM-PETMR showed second and third best performance quantitatively, respectively, but Multi-branch UNet caused oversmoothing and DDPM-PETMR caused uneven boundary. 
CycleWGAN and Pix2Pix reconstructed these regions inconsistently, showing patchy restoration, with some areas losing more signal and others appearing overly bright, while IDDPM failed to predict intensity correctly. 
In contrast, M2Diff preserved both the structural and metabolic distributions, thereby preserving the diagnostically significant features such as hypometabolic regions, and asymmetric uptake patterns. 

\begin{table}[h!]
\caption{Quantitative results for comparative study on ADNI dataset}
\label{table_ADNI}
\centering
\resizebox{\linewidth}{!}{%
\begin{tabular}{lllll}
\toprule
\multicolumn{1}{c}{\textbf{Model}} 
& \multicolumn{1}{c}{\textbf{SSIM $\uparrow$}} 
& \multicolumn{1}{c}{\textbf{PSNR $\uparrow$}} 
& \multicolumn{1}{c}{\textbf{NMSE $\downarrow$}} 
& \multicolumn{1}{c}{\textbf{LPIPS $\downarrow$}} \\
\midrule
CycleWGAN~\cite{zhou2020supervised}     & 0.8949$\pm$\scriptsize0.0376          & 26.3265$\pm$\scriptsize2.0019          & 0.0470$\pm$\scriptsize0.0524          & 0.0442$\pm$\scriptsize0.0161          \\ 
Pix2PixHD~\cite{hosch2022artificial} & 0.9132$\pm$\scriptsize0.0356          & 28.0336$\pm$\scriptsize2.0590          & 0.0344$\pm$\scriptsize0.0482          & 0.0305$\pm$\scriptsize0.0152          \\ 
Multi-Branch UNet~\cite{xie2021anatomically}   &
0.9268$\pm$\scriptsize0.0299   &
28.1862$\pm$\scriptsize2.7187    &   
0.0292$\pm$\scriptsize0.0310    &
0.0623$\pm$\scriptsize0.0413    \\
IDDPM~\cite{yu2025robust}           & 0.9195$\pm$\scriptsize0.0322 & 28.0832$\pm$\scriptsize1.8318 & 0.0334$\pm$\scriptsize0.0444 & \underline{0.0287$\pm$\scriptsize0.0116} \\
DDPM-PETMR~\cite{gong2024pet}                   & \underline{0.9277$\pm$\scriptsize0.0310}                          & \underline{28.7957$\pm$\scriptsize1.9014}                          & \textbf{0.0260$\pm$\scriptsize0.0364} & 
0.0361$\pm$\scriptsize0.0107    \\
M2Diff         & \textbf{0.9370$\pm$\scriptsize0.0310} & \textbf{29.4596$\pm$\scriptsize2.0220} & \underline{0.0266$\pm$\scriptsize0.0412} & \textbf{0.0271$\pm$\scriptsize0.0138} \\ 
\bottomrule
\end{tabular}
}
\end{table}

M2Diff consistently outperformed all the baselines by leveraging anatomical guidance, multi-task learning, and hierarchical feature fusion. Some discrepancies were seen between diffusion and non-diffusion based methods. With the DaCRA dataset, non-diffusion based methods outperformed diffusion based methods likely due to the effectiveness of adversarial training and image-to-image translation on small and low-variability datasets. In contrast, diffusion-based methods, which model a full probability distribution, struggled in this constrained setting. However, with the ADNI dataset, diffusion-based methods surpassed non-diffusion based methods by potentially capturing complex structural and metabolic variations, whereas non-diffusion based methods failed to generalise well to this higher variability.
This observation aligns with prior findings where GANs outperformed diffusion models under specific conditions. For instance, \cite{shi2024diffusion} reported superior performance of U-Net-based models over diffusion approaches in segmentation tasks. Similarly, \cite{muller2024diffusion} found that while GANs performed better for low-dose recovery in breast MRI, they also introduced more false positives during clinical evaluation, a trend we observe in Fig.~\ref{fig_baseline}, where GAN-based models generate additional uptake regions.
In both cases, M2Diff still surpasses baseline models by using its multi-task architecture and feature fusion to preserve features from each modality.

\begin{figure}
\centering
\includegraphics[width=\linewidth]{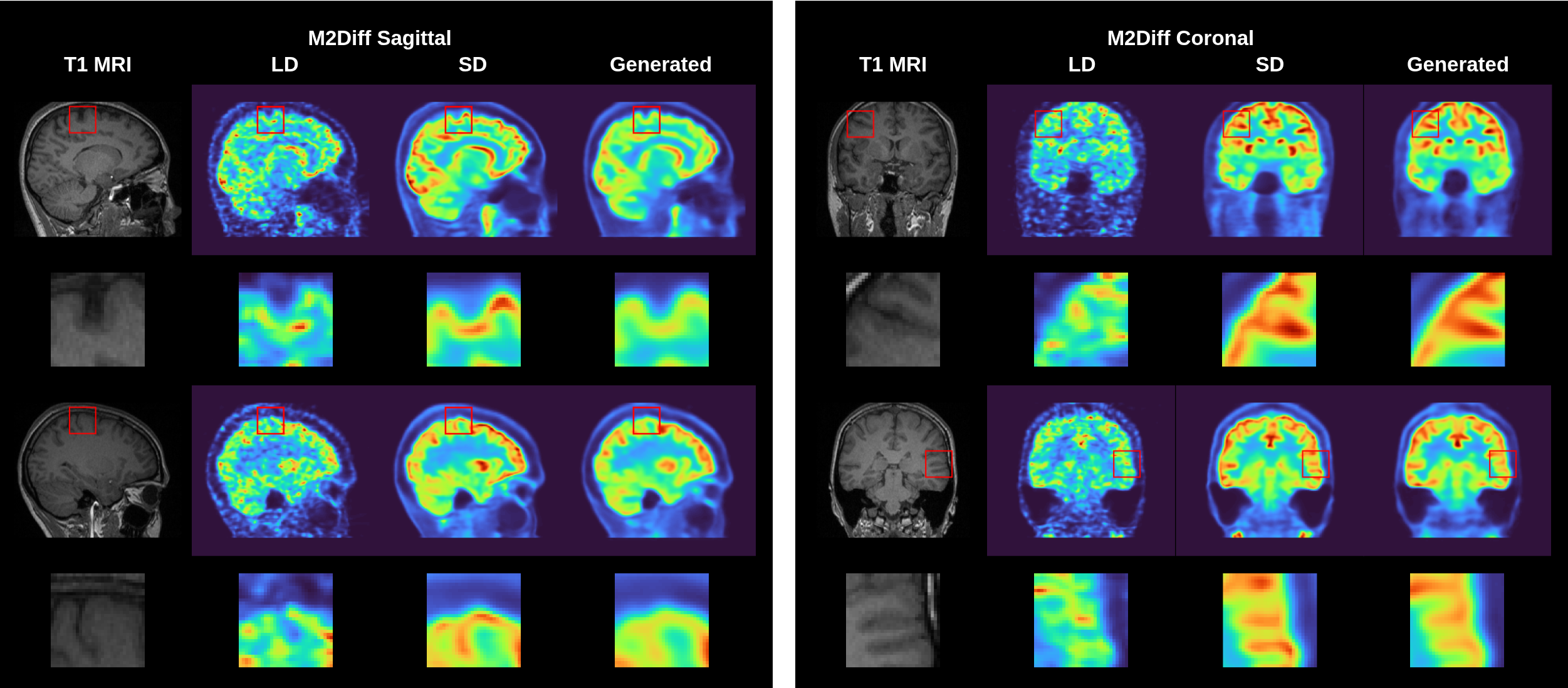}
\caption{Qualitative visualization of the proposed M2Diff model’s performance across reconstructed sagittal and coronal views. Each column shows the corresponding T1-weighted MRI, low-dose (LD) PET input, standard-dose (SD) reference, and reconstructed PET output. The upper
portion of each tile shows the full axial brain slice, while the lower zoomed-in
views (red-bordered) highlight regions of interest. All PET images, including zoomed ROIs, are displayed using a fixed intensity range of [0,1] for all methods.}
\label{fig_sag_cor}
\end{figure}

One limitation observed during visualization was the inconsistency across coronal and sagittal views when reconstructing 3D volumes by stacking 2D axial slices. Although this approach showed promise, slight inter-slice intensity variations and interpolation errors introduced visual artifacts across planes. To address this, we trained and evaluated the M2Diff model on 2D slices extracted along the axial, coronal, and sagittal orientations. These slices were then processed according to the scheme suggested by Lucas \etal~\cite{lucas2025multisequence}. Predicted slices from all 3 orientations were stacked to generate 3 volumes. After applying the Fourier wavelet transform, all 3 volumes were averaged out to get a single 3D volume. The reconstructed slices were then visualised as 2D projections, enabling direct qualitative comparison of anatomical fidelity and tracer uptake patterns across different planes. Fig.~\ref{fig_sag_cor} shows reconstruction results from coronal and sagittal views.

\subsection{Statistical Analysis}

Across all experimental settings, paired t-tests were conducted to evaluate the statistical significance of performance differences between M2Diff and competing baselines. A two-tailed paired t-test was conducted to evaluate statistical differences among methods. Overall, M2Diff consistently demonstrated superior quantitative performance, with highly significant differences across the majority of metrics, confirming the robustness and reliability of its gains.

For the DaCRA dataset with DRF of $\times100$, M2Diff demonstrated significant improvements (p $<$ 0.001) over DiffusionMTL, CDM, and all GAN-based models in SSIM, PSNR, and LPIPS, indicating stronger structural fidelity and perceptual realism. The reduction in NMSE (p $<$ 0.001) was significant when compared with DiffusionMTL, confirming improved voxel-level accuracy. Compared with IDDPM, M2Diff achieved significantly higher SSIM (p $<$ 0.001), PSNR (p $<$ 0.001), and lower LPIPS (p $<$ 0.05), while the NMSE difference (p $>$ 0.05) was not statistically significant, suggesting comparable error variance between the two diffusion models. Similarly, compared to DDPM-PETMR, none of the metric differences reached significance (all p $>$ 0.05), indicating comparable performance when both models leverage diffusion-based denoising. Nevertheless, M2Diff’s advantage in cross-modal fusion and perceptual consistency remained evident when compared with conventional and GAN baselines, where all differences were statistically significant (p $<$ 0.001).

At a DRF of $\times20$, M2Diff consistently outperformed all comparison methods, with nearly all results reaching extreme statistical significance (p $<$ 0.001). SSIM and PSNR gains over DiffusionMTL, DDPM, IDDPM, CycleGAN, and Pix2Pix were highly significant (p $<$ 0.001), demonstrating the method’s consistent superiority across both structural and perceptual domains. Compared with MB-UNet, M2Diff achieved significantly higher PSNR (p $<$ 0.001), lower NMSE (p $<$ 0.001), and improved LPIPS (p $<$ 0.001), though SSIM results were comparable. These results indicate that even under less aggressive noise conditions, M2Diff retains its quantitative edge and maintains statistically significant improvements in reconstruction quality.

On the ADNI dataset, having higher inter-subject variability, M2Diff maintained significant superiority over all baselines. Paired t-tests revealed highly significant improvements (p $<$ 0.001) in SSIM, PSNR, and LPIPS across all comparisons, confirming that the observed performance gains are not due to random variation. The only exception was NMSE versus DDPM-PETMR, where both models performed comparably in terms of absolute error. Compared with IDDPM and MB-UNet, M2Diff showed significant improvements across all metrics (p $<$ 0.001), demonstrating superior structural consistency and perceptual accuracy. Compared with CycleGAN and Pix2Pix, the differences were again statistically significant across all metrics, underscoring M2Diff’s ability to avoid over-smoothing and artifact generation typical of adversarial models. Collectively, these findings confirm that M2Diff achieves statistically validated improvements in both objective fidelity and perceptual quality across datasets and dose levels.

\begin{table*}[h!]
\caption{Quantitative results of ablation study for the proposed model on the DaCRA dataset. \textbf{T1} indicates whether MRI guidance is used; 
\textbf{Task 2} refers to the secondary MRI-to-PET generation pathway; 
\textbf{Encoders} and \textbf{Decoders} specify number of encoders and decoders used in the architecture; 
\textbf{HFF} denotes the presence of the Hierarchical Feature Fusion module; 
\textbf{Asym. Decoders} indicates whether different decoder settings (e.g., dropout) are used for PET and MRI branches.}
\label{table_ablation}
\centering
\resizebox{0.8\linewidth}{!}{%
\begin{tabular}{lccllcccccc}
\hline
\textbf{Version}        &   \textbf{T1}               & \textbf{Task 2}           & \textbf{Encoders} & \textbf{Decoders} & \textbf{HFF}              & \textbf{Asym. Decoders}   & \textbf{SSIM $\uparrow$}                                             & \textbf{PSNR $\uparrow$}                                              & \textbf{NMSE $\downarrow$}                                           & \textbf{LPIPS $\downarrow$}                                    \\ \hline          V1      &            &                           & 1                 & 1                 &                           &                           & 0.9340$\pm$\scriptsize0.0414                          & 27.1968$\pm$\scriptsize2.3031                          & 0.0751$\pm$\scriptsize0.0939                          & 0.0434$\pm$\scriptsize0.0407                    \\      V2      &
\checkmark &                           & 1                 & 1                 &                           &                           & 0.9462$\pm$\scriptsize0.0335                          & 28.1048$\pm$\scriptsize2.5155                          & \underline{0.0646$\pm$\scriptsize0.0766} & 0.0378$\pm$\scriptsize0.0277                    \\     V3      &
\checkmark & \checkmark & 2                 & 2                 &                           &                           & 0.9267$\pm$\scriptsize0.0386                          & 26.5145$\pm$\scriptsize2.2397                          & 0.0931$\pm$\scriptsize0.1183                          & 0.0481$\pm$\scriptsize0.0359                    \\       V4      &
\checkmark & \multicolumn{1}{l}{}      & 2                 & 1                 & \checkmark & \multicolumn{1}{l}{}      & \multicolumn{1}{l}{0.9455$\pm$0.0328}                                & \multicolumn{1}{l}{27.9284$\pm$2.2955}                                & \multicolumn{1}{l}{0.0683$\pm$0.0887}                                & \multicolumn{1}{l}{\underline{0.0360$\pm$0.0293}} \\        V5      &
\checkmark & \checkmark & 2                 & 2                 & \checkmark & \checkmark & \underline{0.9494$\pm$\scriptsize0.0329} & \underline{28.1199$\pm$\scriptsize2.6012} & \textbf{0.0646$\pm$\scriptsize0.0752}                 & 0.0364$\pm$\scriptsize0.0322                    \\       M2Diff      &
\checkmark & \checkmark & 2                 & 2                 & \checkmark &                           & \textbf{0.9528$\pm$\scriptsize0.0308}                 & \textbf{28.6369$\pm$\scriptsize2.7353}                 & 0.0694$\pm$\scriptsize0.1182                          & \textbf{0.0349$\pm$\scriptsize0.0298}           \\ \hline
\end{tabular}
}
\end{table*}

\subsection{Ablation Study}

To verify the effectiveness of M2Diff further, we conducted a series of experiments with the following variants: \textbf{V1:} Single Task M2Diff only conditioned on LD-PET images; \textbf{V2:} Single Task M2Diff conditioned on LD-PET and T1-weighted images; \textbf{V3:} M2Diff, where the first task model is conditioned on LD-PET and the second task model is conditioned on T1-weighted MRI without HFF; \textbf{V4:} M2Diff with one combined decoder with HFF; \textbf{V5:} M2Diff with HFF and asymmetric decoders indicating whether different decoder settings (e.g., dropout) are used for PET and MRI branches; \textbf{V6:} M2Diff with HFF and symmetric decoders, which is the final model. We evaluated this ablation study on the DaCRA dataset.

\begin{figure}[h!]
\centering
\includegraphics[width=\linewidth]{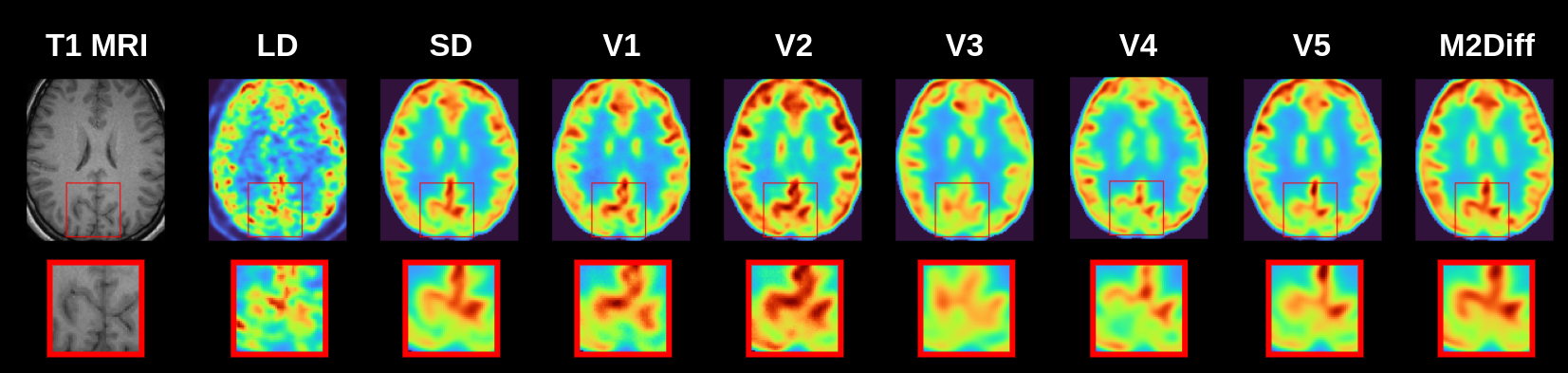}
\caption{Qualitative comparison of reconstructed PET images across different variants of M2Diff as mentioned in Table~\ref{table_ablation} on the DaCRA dataset. Each column represents a different reconstruction method, including T1-weighted MRI, low-dose input (LD), standard-dose ground truth (SD), and different M2Diff variants. The upper portion of each tile shows the full axial brain slice, while the lower zoomed-in views (red-bordered) highlight regions of interest. All PET images, including zoomed ROIs, are displayed using a fixed intensity range of [0,1] for all methods.} 
\label{fig_ablation}
\end{figure}

Qualitative results are shown in Fig.~\ref{fig_ablation} and quantitative results are shown in Table~\ref{table_ablation}. Introducing T1 as a conditional mechanism along with LD-PET has improved the performance of M2Diff when compared to single-task M2Diff, only conditioned on LD-PET. However, our proposed model, which has a multi-task architecture and fusion features, has improved the performance even further. This shows the significance of M2Diff when using multiple modalities in LD-to-SD PET recovery. By looking at qualitative results, it can be seen that our modifications have effectively resulted in the reconstruction of small details. We also showed the effectiveness of HFF by removing it from the M2Diff model. Without feature fusion, both task-specific models worked separately, and the only common factor guiding both tasks was the loss function. This caused a reduction in the model's accuracy, and a significant decrease in the model's performance was observed.
The true advantage of the multi-task model lies in the effective information sharing between tasks. Processing both modalities separately helps in extracting features effectively, but this setup is of no use if these features are not shared at the decoder stage. When HFF is removed, this cross-task communication is essentially lost. In such a setup, the only shared component is the bias loss, which becomes insufficient and potentially misleading. Without shared features, the supervision from the auxiliary task lacks grounding in relevant representations, causing both task branches to diverge or underperform. This explains why M2Diff (w/o HFF), despite retaining the multi-task setup, performs worse than even the simpler single-task variant (M2Diff w/o T1 \& Task 2).

To further investigate the design choices and the impact of decoder-level regularization on model performance, we evaluated two configurations: one employing a single decoder with hierarchical feature fusion (HFF), and the other using asymmetric dual decoders with non-identical dropout settings. The overall architecture was kept fixed, while dropout was selectively varied between the PET and MRI branches. Our results revealed a performance decline in both configurations, emphasizing that identical decoder setups yield more stable and consistent performance, thereby suggesting that a symmetric decoder design facilitates better feature alignment during multi-modal fusion.

\subsection{MRI-free Inference}

Recognizing that MRI data are not always available alongside PET scans, we trained our model to operate effectively without full MRI guidance. Inspired by classifier-free guidance~\cite{ho2022classifier}, we proposed a training strategy that introduces partial MRI conditioning during training. A control flag was implemented to indicate whether MRI data were available for a given sample; when set to 0, the model trained the PET pathway independently, without incorporating MRI features. This strategy not only enhanced flexibility across different data-availability scenarios but also reduced computational complexity during inference for PET-only configurations.  

\begin{figure}[h!]
\centering
\includegraphics[width=\linewidth]{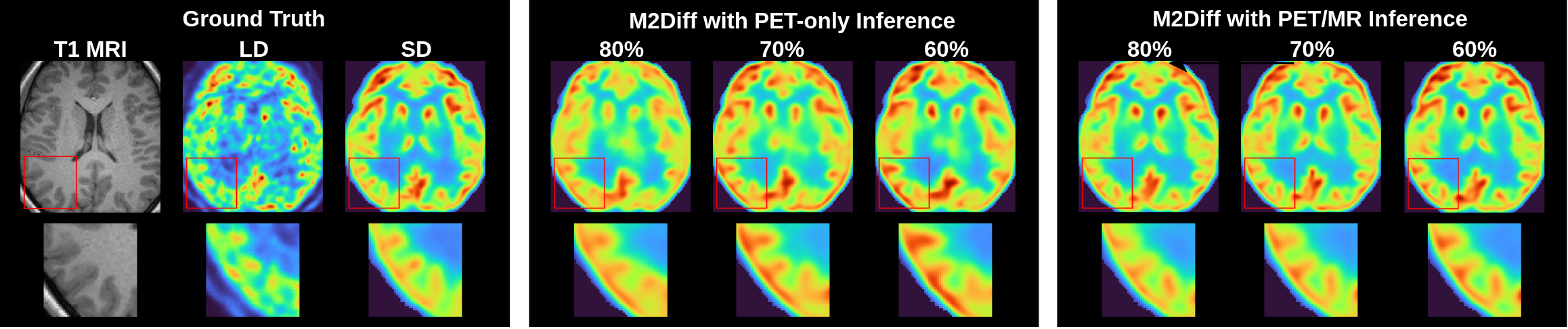}
\caption{Qualitative comparison of M2Diff performance with and without MRI guidance under varying training data proportions. The first three columns show the reference modalities: T1-weighted MRI, low-dose (LD) PET, and standard-dose (SD) PET. The subsequent columns depict (i) M2Diff PET-only after re-training for 80\%, 70\%, and 60\% of the training with MRI, and (ii) M2Diff PET/MR after re-training at the same data proportions during training. Zoomed-in regions highlight cortical structures. All PET images, including zoomed ROIs, are displayed using a fixed intensity range of [0,1] for all methods.}
\label{fig_ablation_wo_mri}
\end{figure}

Furthermore, we evaluated model performance under varying levels of MRI availability during training on the DaCRA dataset. Qualitative and quantitative comparisons are presented in Fig.~\ref{fig_ablation_wo_mri} and Table~\ref{table_ablation_wo_mri}, respectively. The results show that when MRI data are available at test time, the model trained with partial MRI conditioning performs comparably to one trained with full (100\%) MRI guidance. Conversely, when MRI data are absent, it surpasses the IDDPM baseline trained solely on PET data. An additional ablation study varying MRI availability to 80\%, 70\%, and 60\% of the training samples revealed that the best results were achieved when MRI guidance was provided in 70\% of cases.

\begin{table}[h!]
\caption{Quantitative evaluation of M2Diff on DaCRA dataset trained with varying proportions of MRI-guided data (60–80\%) and tested with or without MRI guidance.}
\label{table_ablation_wo_mri}
\resizebox{\linewidth}{!}{%
\begin{tabular}{cccccc}
\hline
\textbf{MRI Usage} & \textbf{Testing Mode} & \textbf{SSIM $\uparrow$} & \textbf{PSNR $\uparrow$} & \textbf{NMSE $\downarrow$} & \textbf{LPIPS $\downarrow$} \\
\hline
\multirow{2}{*}{80\%} & With MRI    & 0.9483$\pm$\scriptsize0.0316 & 28.4979$\pm$\scriptsize2.6567 & 0.0699$\pm$\scriptsize0.1129 & 0.0375$\pm$\scriptsize0.0285 \\
 & Without MRI & 0.9261$\pm$\scriptsize0.0345 & 26.6807$\pm$\scriptsize2.0517 & 0.0993$\pm$\scriptsize0.1690 & 0.0555$\pm$\scriptsize0.0318 \\
\hline
\multirow{2}{*}{70\%} & With MRI    & 0.9531$\pm$\scriptsize0.0308 & 28.8860$\pm$\scriptsize2.6141 & 0.0615$\pm$\scriptsize0.0932 & 0.0356$\pm$\scriptsize0.0288 \\
 & Without MRI & 0.9363$\pm$\scriptsize0.0340 & 27.3164$\pm$\scriptsize2.1751 & 0.0846$\pm$\scriptsize0.1296 & 0.0481$\pm$\scriptsize0.0330 \\
\hline
\multirow{2}{*}{60\%} & With MRI    & 0.9519$\pm$\scriptsize0.0330 & 28.7944$\pm$\scriptsize2.7035 & 0.0611$\pm$\scriptsize0.0846 & 0.0367$\pm$\scriptsize0.0335 \\
 & Without MRI & 0.9342$\pm$\scriptsize0.0374 & 27.0859$\pm$\scriptsize2.1341 & 0.0866$\pm$\scriptsize0.1260 & 0.0485$\pm$\scriptsize0.0358 \\
\hline
\end{tabular}}
\end{table}

\subsection{Computational Complexity Analysis}

The computational and quantitative analysis highlights the trade-off between model complexity, reconstruction fidelity, and runtime efficiency. Here, we report the training and testing times per epoch (batch size = 1) and per subject, respectively, on the DaCRA dataset with an image size of 256×256, using an NVIDIA A40 GPU. The compact MB-UNet model, with only 2.90M parameters, achieved the fastest training speed of 218.28s per epoch and 0.01s of inference time per slice, demonstrating its suitability for lightweight applications, though at the cost of lower perceptual and overall quality on datasets with complex features. The IDDPM and IDDPM PET-MR models, each comprising 124.05M parameters, substantially improved PET reconstruction quality by better capturing noise statistics and anatomical structure. Their training times were 429.46s and 489.22s per epoch with a batch size of 1, respectively, while the testing time was $\sim$40s per subject. The DiffusionMTL framework, with 231.48M parameters, further incorporated multi-task learning at the expense of greater computational demand, taking 1056.46s per epoch with a batch size of 1 and $\sim$73s per subject during inference. Despite being computationally intensive, this model did not perform well, indicating that our architectural choices make M2Diff perform better, not the high number of parameters. The proposed M2Diff model, integrating dual decoders and hierarchical feature fusion, had the largest capacity, with 268.99M parameters, and required 1217.84s per epoch with a batch size of 1 and approximately 88s to reconstruct a full PET subject. Despite this increased runtime, M2Diff achieved the best quantitative performance, markedly improving SSIM and PSNR. Overall, these results demonstrate that M2Diff offers the optimal balance between reconstruction accuracy and computational feasibility, achieving superior multimodal restoration while maintaining practical inference times at the subject level.

\subsection{CKA-Based Representation Similarity Analysis}

To investigate how modality-specific representation evolves within the network, we performed Centered Kernel Alignment (CKA)~\cite{pmlr-v97-kornblith19a} between PET and MRI conditioned branches for the encoder and decoder networks in a multi-task model.
CKA is a similarity metric used to compare the representations learned by different neural network layers or models using the Hilbert-Schmidt Independence Criterion (HSIC)~\cite{wang2021learning}. 

\begin{figure}[h!]
    \centering
    \includegraphics[width=1.0\linewidth]{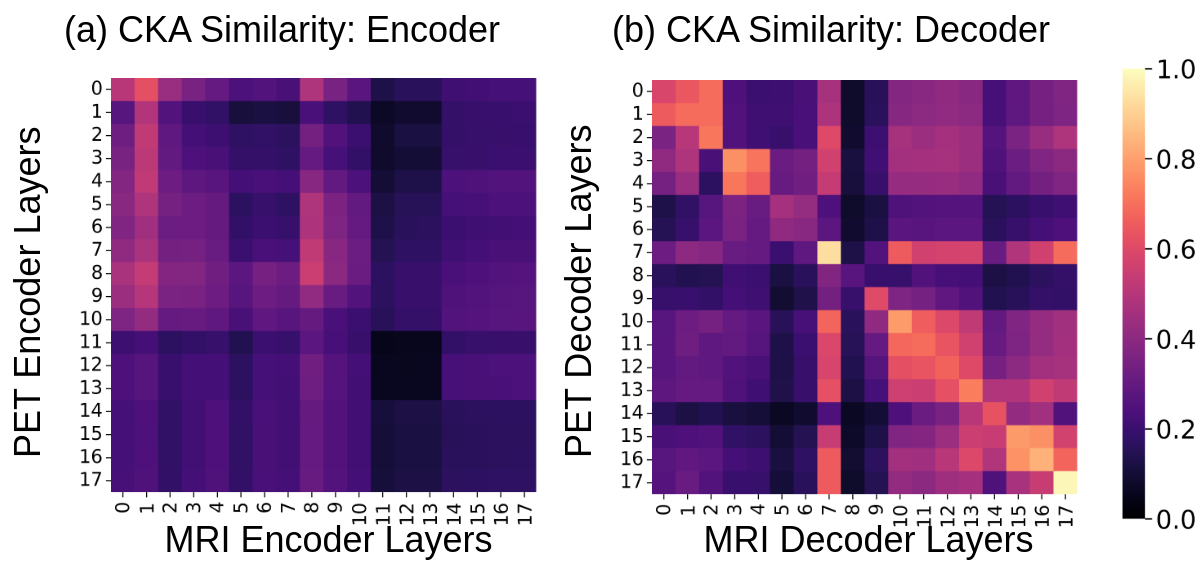}
    \caption{Centered Kernel Alignment (CKA) similarity matrices between PET and MRI branches. Here, CKA = 0 and 1 indicate that the two layers are dissimilar and similar, respectively. (a) Encoder layers. (b) Decoder layers.}
    \label{fig:cka_mat}
\end{figure}

Fig.~\ref{fig:cka_mat}(a) shows a similarity matrix for the encoder layers for the PET and MRI conditioned branches. 
High CKA in shallow layers (~0.62) suggests shared low-level features, while lower values in deeper layers indicate learning of modality-specific representations. In contrast, Fig.~\ref{fig:cka_mat}(b) shows CKA for decoder layers where a different trend can be observed. Decoder layers show increasing CKA, peaking at 0.99 in the final layer, indicating convergence to shared representations for SD-PET reconstruction despite separate latent paths.

These findings validate our multi-task design, where the encoder extracts distinct modality-specific features, and the decoder aligns them for a unified output. Low encoder CKA scores support disentangled representation learning, while high decoder CKA scores confirm effective fusion and shared supervision.

\section{Conclusion}

In this paper, we introduce M2Diff, a multi-modality multi-task IDDPM for LD-to-SD PET recovery. By leveraging a dual-encoder architecture, this model captures modality-specific features and avoids early feature fusion. A hierarchical feature fusion mechanism enables layer-wise integration of complementary information and feature sharing among both tasks at the decoder stage, resulting in high-fidelity SD-PET image generation. This architecture promotes disentangled encoding and shared decoding, a balance that is both principled and empirically effective.

A comprehensive evaluation is performed using the DaCRA and ADNI datasets, where M2Diff outperforms baseline methods, including IDDPM and other state-of-the-art methods, based on both quantitative metrics (PSNR, NMSE, SSIM, LPIP) and qualitative assessment. Notably, the recovered PET scans demonstrate improved preservation of anatomical structure and metabolic distribution, key indicators of diagnostic utility. Additionally, the paired t-test indicates a statistically significant improvement in performance for M2Diff. These results validate the M2Diff for reconstructing clinically relevant features, such as regional hypometabolism and disease-specific uptake patterns, which are often degraded in low-dose settings. The ablation study of the model further sheds light on the importance of the multi-task architecture and the hierarchical feature fusion block. Furthermore, the CKA-based similarity matrix provides insights into the model's internal representations, revealing that encoder branches learns distinct task-specific features and decoder layers converge to shared representations aligned with the reconstruction objective. These findings support the hypothesis that modality-specific encoding paired with unified decoding is a powerful paradigm for multi-modal medical image generation.

Despite these promising results, there are several limitations that need to be addressed. First, the current model operates on 2D, which, while effective as a proof of concept, does not fully capture the spatial continuity and volumetric context present in 3D PET imaging. Secondly, the approach relies on paired PET-MR data, which limits the applicability where multi-model data is not available. Thirdly, the model has not yet been further evaluated for clinical diagnostic utility by physicians, which is critical for translational adoption. To mitigate the first two limitations, we implemented a partial MRI training strategy with weight transfer to a PET-only pipeline and reconstructed 3D volumes from multi-orientation (axial, coronal, sagittal) 2D slices, providing a practical compromise between data requirements and spatial consistency.

Future work will aim to address these limitations by extending the model to a fully 3D framework to better exploit spatial dependencies, adopting an unpaired or weakly supervised framework, and performing clinical validation studies. These steps will help broaden the impact of M2Diff and contribute towards safer, faster, and more accurate PET imaging in a real-world clinical environment.

\section*{Acknowledgments}
The authors would like to thank the collaborators and colleagues for helpful discussions and technical support during this research. This work was supported in part by the Australian Research Council, and Zhaolin Chen is supported by an Australian Research Council Fellowship Grant \#IM230100002.

\section*{Conflict of Interest}
The authors declare that they have no conflicts of interest related to this work.
 
%

\bibliographystyle{IEEEtran}
\bibliography{references}

\end{document}